\documentclass[%
 reprint,
 amsmath,amssymb,
 aps,
]{revtex4-1}
\usepackage{afterpage}
\usepackage{color}
\usepackage{graphicx}
\usepackage{dcolumn}
\usepackage{bm}

\newcommand{\comment}[1]{}

\begin{document}

\preprint{APS/123-QED}

\title{Quantum Spin Superfluid from Bose-Einstein Condensation of Spinons in Pyrochlore Spin Ice}

\author{Imam Makhfudz}
\affiliation{%
Univ Lyon, ENS de Lyon, Univ Claude Bernard, \comment{CNRS, Laboratoire de Physique, F-69342} Lyon, France\\ 
}%
\date{\today}

\begin{abstract}
Quantum spin ice in pyrochlore lattice exemplifies three dimensional frustrated spin systems.In existing studies, Bose-Einstein condensation of bosonic spinons gives rise to magnetically ordered ground state.A truly liquid quantum spin superfluid state that manifests a deconfined Higgs condensate of spinons is demonstrated.This state is shown to occur in the fully antiferromagnetic case of the non-Kramers quantum spin ice with a fluctuations-induced first-order quantum phase transition from the $U(1)$ spin liquid to the spin superfluid.The spin superfluid density jump is obtained analytically.




\end{abstract}

\pacs{Valid PACS appear here}
\maketitle


\textit{Introduction.\textemdash}
Superfluidity discovered in the 30's \cite{Kapitza} gives an example of manifestation of quantum mechanics at macroscopic level and for decades has fascinated physicists not only in low-temperature physics community but also nowadays has inspired new ideas based on it in such diverse fields as astrophysics and cosmology.Its theoretical explanation in liquid Helium 4 \cite{Tisza}\cite{Landau}\cite{Ginzburg} and liquid Helium 3 \cite{Leggett} constitutes a crowning achievement of theoretical condensed matter physics in the 20$^{\mathrm{th}}$ century, accompanying that of BCS theory of superconductivity \cite{BCS}.In the context of magnetism, the interplay between geometric frustration and quantum fluctuations can give rise to a disordered spin state even at $T=0$, which would otherwise be forbidden by classical physics.This quantum spin liquid (QSL) state was first suggested by Phil Anderson in terms of resonating valence bond state in 1973 \cite{AndersonRVB} and also made appearance in the theories of high $T_c$ superconductivity in copper oxides.Essentially, quantum spin liquid is a liquid state of spins, which has the maximum possible symmetry.In spin systems with strong quantum fluctuations, enhanced by lattice geometry such as in frustrated spin systems \cite{BalentsNature}, such quantum spin disordered state is prevalent.Fractionalization of the spin into emergent quantum particles and their deconfinement occur in quantum spin liquid \cite{Balents}.In 3d frustrated spin systems such as those forming a family called pyrochlore quantum spin ice where a $U(1)$ spin liquid is argued to exist, the emergent particle manifests as bosonic spinon that act like magnetic monopoles \cite{Hermele}.Bose-Einstein condensation of such bosons which manifests Higgs mechanism in confined state has so far been found to give rise to some form of magnetic order \cite{SavaryBalents}\cite{LeeOnodaBalents}\cite{GangChen}, as is the case also for the condensation of magnons \cite{MatsubaraMatsuda}\cite{Oshikawa}\cite{GiamarchiRueggTchernyshyov}.

In this work, we propose a different scenario where Bose-Einstein condensation of bosonic spinons, which also manifests a Higgs mechanism but in deconfined state, gives rise to a spin analog of the superfluid in liquid Helium, which we refer to as quantum spin superfluid state.This article will show how such exotic state emerges in the ground state of pyrochlore spin ice.This state preserves the translational and spin rotational symmetries and thus represents a liquid state of spins but yet it has nonzero spin superfluid density, due to the breaking of $U(1)$ boson number gauge invariance.Using variational energy calculation, we deduce that the quantum spin superfluid occurs next to the $U(1)$ QSL regime of the fully antiferromagnetic non-Kramers doublet quantum spin ice.The $U(1)$ QSL-quantum spin superfluid transition is shown clearly to be first order, marked by a finite spin superfluid density jump which we derive analytically.

\textit{Generic Model: Bosonic Spinon-Gauge Field Hamiltonian.\textemdash}
We adopt a bosonic formalism applied to pyrochlore quantum spin ice and apply the so-called gauge mean-field theory (gMFT), which is a mean-field theory that incorporates the gauge degree of freedom and thus takes into account the gauge (quantum) fluctuations in the ground state of quantum spin ice \cite{SavaryBalents}.When implemented to generic quantum spin ice systems of non-Kramers doublet with integer spins, one obtains an effective Hamiltonian in terms of bosonic spinon and the gauge field operator, which takes the form of lattice quantum electrodynamics (QED) \cite{LeeOnodaBalents}.Using simpler but otherwise equivalent notation to that in \cite{LeeOnodaBalents}, the bosonic QED model of quantum spin ice is given by
\[
H_{\mathrm{QED}}=\frac{J_{zz}}{2}\sum_{\mathbf{r}}Q^2_{\mathbf{r}}-J_{\pm}\sum_{\mathbf{r}}\sum_{\mu\neq\nu}b^{\dag}_{\mathbf{r}_{\mu}}b_{\mathbf{r}_{\nu}}s^{-\eta_{\mathbf{r}}}_{\mathbf{r}\mathbf{r}_{\mu}}s^{+\eta_{\mathbf{r}}}_{\mathbf{r}\mathbf{r}_{\nu}}
\]
\[
+\frac{J_{\pm\pm}}{2}\sum_{\mathbf{r}}\sum_{\mu\neq\nu}\left(\gamma^{-2\eta_r}_{\mu\nu}b^{\dag}_{\mathbf{r}}b^{\dag}_{\mathbf{r}}b_{\mathbf{r}_{\mu}}b_{\mathbf{r}_{\nu}}s^{\eta_{\mathbf{r}}}_{\mathbf{r}\mathbf{r}_{\mu}}s^{\eta_{\mathbf{r}}}_{\mathbf{r}\mathbf{r}_{\nu}}+\mathrm{h.c.}\right)
\]
\begin{equation}\label{latticeQED}
-J_{z\pm}\sum_{\mathbf{r}}\sum_{\mu\neq\nu}S^{z}_{\mathbf{r}\mathbf{r}_{\mu}}\left(\gamma^{-\eta_r}_{\mu\nu}b^{\dag}_{\mathbf{r}}b_{\mathbf{r}_{\nu}}s^{\eta_{\mathbf{r}}}_{\mathbf{r}\mathbf{r}_{\nu}}+\mathrm{h.c.}\right)+\mathrm{constant}
\end{equation}
where $b^{\dag}_{\mathbf{r}}(b_{\mathbf{r}})$ is bosonic spinon creation (annihilation) operator defined at the position vector $\mathbf{r}$ of the center of a tetrahedron of the pyrochlore lattice, $\mathbf{r}_{\mu}=\mathbf{r}+\eta_{\mathbf{r}}\mathbf{e}_{\mu}$ where $\mathbf{e}_{\mu}$ with $\eta_{\mathbf{r}}=\pm 1$ for the `up' and `down' tetrahedra respectively is the \textit{local} basis vector defined at each corner of the tetrahedron.The spin operators are defined by $S^{\pm}_{\tilde{\mathbf{r}}_{\mu}}=b^{\dag}_{\mathbf{r}}s^{\pm}_{\mathbf{r}\mathbf{r}_{\mu}}b_{\mathbf{r}_{\mu}},S^z_{\tilde{\mathbf{r}}_{\mu}}=s^z_{\mathbf{r}\mathbf{r}_{\mu}}$ where $\tilde{\mathbf{r}}_{\mu}=\mathbf{r}+\eta_{\mathbf{r}}\mathbf{e}_{\mu}/2$ marks the pyrochlore lattice sites, and $\gamma_{\mu\nu}$ are constants \cite{LeeOnodaBalents}.The boson charge $Q_{\mathbf{r}}=\eta_{\mathbf{r}}\sum^3_{\mu=0}S^z_{\mathbf{r},\mathbf{r}+\eta_{\mathbf{r}}\mathbf{e}_{\mu}/2}$ is subject to the constraint that the total electric charge $Q=\sum_{\mathbf{r}}Q_{\mathbf{r}}$ commutes with the Hamiltonian $H_{\mathrm{QED}}$.In existing quantum spin ice materials, $J_{zz},J_{\pm\pm}>0$, $J_{z\pm}<0$ while $J_{\pm}>0$ (ferromagnetic) or $J_{\pm}<0$ (antiferromagnetic) \cite{RossPRX}.

\textit{`Unilinear' Gauge Mean-Field Theory:The Spin Superfluid Density and Mean-Field Ansatz.\textemdash}
From Eq.(\ref{latticeQED}), we perform a mean-field decoupling that manifests the creation or annihilation of bosonic particles \cite{SachdevQPTbook}, to be called `unilinear mean-field decoupling' here, where we have single boson creation or annihilation operators rather than bilinear products of them in the resulting mean-field Hamiltonian \cite{SuppMat}.Upon the mean-field decomposition, the boson and the gauge field are decoupled and the resulting mean-field QED Hamiltonian can be written as \cite{SuppMat}
\begin{equation}
H^{\mathrm{MF}}_{\mathrm{QED}}=H^{\mathrm{MF}}_{\mathrm{QED}}(b)+H^{\mathrm{MF}}_{\mathrm{QED}}(s^{\pm},s^z)
\end{equation}
where 
\[
H^{\mathrm{MF}}_{\mathrm{QED}}(b)=-J_{\pm}\sum_{\mathbf{r},\mu\neq\nu}g_{-\mu}g_{+\nu}(\Psi b^{\dag}_{\mathbf{r}_{\mu}}+\Psi^*b_{\mathbf{r}_{\nu}})+\frac{3J_{\pm\pm}}{2}\times
\]
\[
\sum_{\mathbf{r}, \mu\neq\nu}|\Psi|^2\left[\gamma^{-2\eta_{\mathbf{r}}}_{\mu\nu}g_{+\mu}g_{+\nu}\left(2\Psi b^{\dag}_{\mathbf{r}}+\Psi^*(b_{\mathbf{r}_{\mu}}+b_{\mathbf{r}_{\nu}})\right)+\mathrm{h.c.}\right]
\]
\begin{equation}\label{HMFQEDboson}
-J_{z\pm}\sum_{\mathbf{r},\mu\neq\nu}\left(\gamma^{-\eta_{\mathbf{r}}}_{\mu\nu}g_{z\mu}g_{+\nu} (\Psi b^{\dag}_{\mathbf{r}}+\Psi^* b_{\mathbf{r}_{\nu}})+\mathrm{h.c.}\right)
\end{equation}
\[
H^{\mathrm{MF}}_{\mathrm{QED}}(s^{\pm},s^z)=
\frac{J_{zz}}{2}\sum_{\mathbf{r},\mu,\nu}\left(s^z_{\mathbf{r}\mathbf{r}_{\mu}}g_{z\nu}+s^z_{\mathbf{r}\mathbf{r}_{\nu}}g_{z\mu}\right)
\]
\[
-J_{\pm}\sum_{\mathbf{r},\mu\neq\nu}|\Psi|^2\left(s^{+\eta_{\mathbf{r}}}_{\mathbf{r}\mathbf{r}_{\nu}}g_{-\mu}+g_{+\nu} s^{-\eta_{\mathbf{r}}}_{\mathbf{r}\mathbf{r}_{\mu}}\right)
\]
\[
+\frac{3}{2}J_{\pm\pm}\sum_{\mathbf{r},\mu\neq\nu}\left[\gamma^{-2\eta_{\mathbf{r}}}_{\mu\nu}|\Psi|^4(s^{\eta_{\mathbf{r}}}_{\mathbf{r}\mathbf{r}_{\mu}}g_{+\nu}+s^{\eta_{\mathbf{r}}}_{\mathbf{r}\mathbf{r}_{\nu}}g_{+\mu})+\mathrm{h.c.}\right]
\]
\begin{equation}\label{gaugefieldMFhamiltonian}
-J_{z\pm}\sum_{\mathbf{r},\mu\neq \nu}\left[\gamma^{-\eta_{\mathbf{r}}}_{\mu\nu}|\Psi|^2(s^z_{\mathbf{r}\mathbf{r}_{\mu}}g_{+\nu}+g_{z\mu} s^{\eta_{\mathbf{r}}}_{\mathbf{r}\mathbf{r}_{\nu}})+\mathrm{h.c.} \right]
\end{equation}
plus constant, with $g_{\pm\mu}=\langle s^{\pm \eta_{\mathbf{r}}}_{\mathbf{r}\mathbf{r}_{\mu}}\rangle, g_{z\mu}=\langle s^{z}_{\mathbf{r}\mathbf{r}_{\mu}}\rangle$ and $\Psi=\langle b_{\mathbf{r}}\rangle$.

In order to characterize the bosonic spinon condensate, we evaluate $\Psi=|\Psi|e^{i\theta}$.On the other hand, a liquid state of spins by definition preserves translational symmetry, which amounts to spatially uniform mean field solution for $|\Psi|$.Taking $\delta \langle H^{\mathrm{MF}}_{\mathrm{QED}}(b)\rangle/\delta \langle b_{\mathbf{r}_{\nu}}\rangle=0$ gives the following result
\begin{equation}\label{superfluidOP}
|\Psi|^2=\frac{\sum_{\mathbf{r},\mu\neq \nu}J_{\pm}g_{-\mu}g_{+\nu}e^{-i\theta}+2J_{z\pm}\mathrm{Re}[\gamma^{-\eta_{\mathbf{r}}}_{\mu\nu}g_{z\mu} g_{+\nu}e^{-i\theta}]}{3J_{\pm\pm}\sum_{\mathbf{r},\mu\neq\nu}\mathrm{Re}\left[\gamma^{-2\eta_{\mathbf{r}}}_{\mu\nu}g_{+\mu}g_{+\nu}e^{-i\theta}\right]}
\end{equation}
where $\mathrm{Re[\cdots]}$ refers to the real part of $\cdots$ \cite{SuppMat}.Intuitively speaking, $g_{\pm\mu}$ is a measure of quantum fluctuations of the spin at a corner of a tetrahedron of the pyrochlore lattice.Therefore, in the Ising limit $g_{\pm\nu}=0$ and in any other cases one expects $g_{\pm\nu}\neq 0$.We will interpret $|\Psi|^2$ as the spin superfluid density and show soon that it gives a mean-field ansatz which results in proper superfluid properties.Clearly, only if the right hand side gives a real positive number do we have a superfluid state while otherwise we have a $U(1)$ QSL state.From Eq.(\ref{superfluidOP}) in conjunction with Eq.(\ref{latticeQED}), we can conclude that the spin superfluid state is driven by the hopping terms $J_{\pm},J_{z\pm}$ while being inhibited by the repulsive boson interaction term $J_{\pm\pm}$.This is totally consistent with the theoretical understanding of superfluidity in liquid Helium and interacting boson system in general where the mobility of the bosons favors superfluidity while a repulsive interaction between them inhibits this exotic state.

We will assume that the superfluid state does not break Ising symmetry, $g_{z\mu} = 0$, which will be shown later to be consistent.Eq.(\ref{superfluidOP}) then simplifies to be
\begin{equation}\label{quantumspinsuperfluiddensity}
|\Psi|^2=\frac{1}{3}\frac{J_{\pm}}{J_{\pm\pm}}\frac{\sum_{\mathbf{r},\mu\neq\nu}g_{-\mu}g_{+\nu}e^{-i\theta}}{\sum_{\mathbf{r},\mu\neq\nu}\mathrm{Re}\left[\gamma^{-2\eta_{\mathbf{r}}}_{\mu\nu}g_{+\mu}g_{+\nu}e^{-i\theta}\right]}
\end{equation}
which works for both ferromagnetic ($J_{\pm}>0$) and antiferromagnetic ($J_{\pm}<0$) case \cite{SuppMat}.Eq.(\ref{quantumspinsuperfluiddensity}) assumes translational symmetry and thus applies only to distinguishing `normal' spin liquid ($|\Psi|=0$) and superfluid spin liquid ($|\Psi|\neq 0$).The mean-field ansatz for a quantum spin superfluid state is obtained by requiring each of the terms under the sum $\sum_{\mu\neq\nu}$ on the numerator of the right hand side of Eq.(\ref{quantumspinsuperfluiddensity}) to be real in order to give a real value for $|\Psi|$.We write $g_{\pm\mu}=\Delta\exp(\pm i\overline{A}_{\mathbf{r},\mathbf{r}_{\mu}})$ which, using the fact that $H^{\mathrm{MF}}_{\mathrm{QED}}(s^{\pm},s^z)$ is a spin $S=1/2$ operator and our assumption $\langle s^z\rangle=0$, imposes $\Delta=1/2$.We find that a nontrivial ansatz which gives $|\Psi^2|>0$ is given by  
\begin{equation}\label{nontrivialphase}
\overline{A}_{\mathbf{r},\mathbf{r}_{0}}=-\pi/2,\overline{A}_{\mathbf{r},\mathbf{r}_{1}}=\pi/2,
\overline{A}_{\mathbf{r},\mathbf{r}_{2}}=-\pi/2,\overline{A}_{\mathbf{r},\mathbf{r}_{3}}=\pi/2
\end{equation}
with $J_{\pm}<0$ (antiferromagnetic case) \cite{SuppMat}.The boson phase field is given by \cite{SuppMat}
\begin{equation}\label{phasefield}
\theta(\mathbf{r})=\mathbf{G}\cdot\mathbf{r}
\end{equation}
where $\mathbf{G}$ is the reciprocal lattice wave vector of the dual diamond lattice, giving $\theta(\mathbf{r}_{\mu})=2m\pi$, and $m=0,\pm 1,\pm 2,\pm 3,\cdots$.Using the above ansatz, we obtain 
\begin{equation}\label{SFdensity}
|\Psi|^2=-\frac{J_{\pm}}{3J_{\pm\pm}}
\end{equation}
for $J_{\pm}<0$.The corresponding gauge flux $\Phi$ through the hexagon of the dual diamond lattice $\Phi=\nabla\times \overline{A}=\sum_{\mathbf{r}\mathbf{r}'\in\mathrm{hex}}\overline{A}_{\mathbf{r},\mathbf{r}'}$ is $\Phi=\pi$ using Eq.(\ref{nontrivialphase}).The resulting gauge flux ensures that the mean-field ansatz preserves the symmetries of the effective Hamiltonian Eq.(\ref{latticeQED}).With the net $\pi$ flux, the spin superfluid state specified by the ansatz Eq.(\ref{nontrivialphase}) preserves time reversal symmetry.Alternative ansatz that works for the ferromagnetic case ($J_{\pm}>0$) is exemplified by $\overline{A}_{\mathbf{r},\mathbf{r}_{0}}=0,\overline{A}_{\mathbf{r},\mathbf{r}_{1}}=\pi, \overline{A}_{\mathbf{r},\mathbf{r}_{2}}=0,\overline{A}_{\mathbf{r},\mathbf{r}_{3}}=\pi$, corresponding to $\Phi=0$ flux state but this turns out to be energetically unfavorable.

Next, we solve $\delta \langle H^{\mathrm{MF}}_{\mathrm{QED}}(s^{\pm},s^z)\rangle /\delta \langle s^{\pm}\rangle=0$ for $\langle s^{\pm}\rangle$.Our ansatz $\langle s^{\pm \eta_{\mathbf{r}}}_{\mathbf{r}\mathbf{r}_{\mu}}\rangle=(1/2)\exp(\pm i\overline{A}_{\mathbf{r},\mathbf{r}_{\mu}})$ with $\overline{A}_{\mathbf{r},\mathbf{r}_{\mu}}$ given by Eq.(\ref{nontrivialphase}) does satisfy this extremum condition $\delta \langle H^{\mathrm{MF}}_{\mathrm{QED}}(s^{\pm},s^z)\rangle /\delta \langle s^{\pm}\rangle=0$ \cite{SuppMat}.Noting the complexity of Eq.(\ref{gaugefieldMFhamiltonian}), the solution of the extremum condition for $\langle s^{\pm} \rangle$ is complex-valued in general.We suggest that a complex value for $\langle s^{\pm} \rangle$ is an indicator for a liquid state of spins, in agreement also with the complex ansatz for $U(1)$ QSL where $\langle s^{\pm \eta_{\mathbf{r}}}_{\mathbf{r}\mathbf{r}_{\mu}}\rangle=\Delta\exp(\pm i\overline{A}_{\mathbf{r},\mathbf{r}_{\mu}})$ with $\Delta=1/2,\overline{A}_{\mathbf{r},\mathbf{r}_{\mu}}=\epsilon_{\mu}\mathbf{Q}\cdot\mathbf{r}$ with $\epsilon_{\mu}=(0110),\mathbf{Q}=2\pi(100)$,$\mathbf{r}=(n_1,n_2,n_3)/4$ \cite{LeeOnodaBalents}, which however gives $|\Psi|^2=0$ when substituted into Eq.(\ref{quantumspinsuperfluiddensity}).This is to be compared with a real-valued ansatz for $\langle s^{\pm} \rangle$ which turns out to give rise to some type of (quadrupolar) magnetically-ordered state \cite{LeeOnodaBalents}.This complex-valued $\langle s^{\pm} \rangle$ is in agreement with our explicit results in Eq.(\ref{nontrivialphase}) above for a liquid state of spins.

Finally, we solve $\delta \langle H^{\mathrm{MF}}_{\mathrm{QED}}(s^{\pm},s^z)\rangle/\delta \langle s^z\rangle=0$ for $\langle s^z\rangle$.We find that $\langle s^z\rangle =0$ is indeed the exact solution of the mean-field equation when $\overline{A}_{\mathbf{r},\mathbf{r}_{\mu}}$ is given by Eq.(\ref{nontrivialphase}) \cite{SuppMat}.Thus, our quantum spin superfluid state breaks no time reversal or Ising symmetry, $\langle s^z\rangle = 0$, consistent with our earlier assumption used in obtaining Eq.(\ref{quantumspinsuperfluiddensity}).In addition, the superfluid state corresponds to $\Phi=\pi$ flux state, according to the conclusion in the previous paragraph.This matches with the $\pi$-flux state realized in the antiferromagnetic case ($J_{\pm}<0$) \cite{LeeOnodaBalents}, which suggests that the spin superfluid state is realized only in antiferromagnetic case, where indeed the frustration takes effect and broadens the regime of the stability of the liquid state of spins \cite{LeeOnodaBalents}.As noted earlier, the mean-field ansatz in this case has a gauge flux $\Phi=\pi$ through the hexagon of the dual diamond lattice, which suggests a complex ground state flux pattern expected of a frustrated spin system with liquid character for its spin state.

\begin{figure}
 \includegraphics[angle=0,origin=c, scale=0.20]{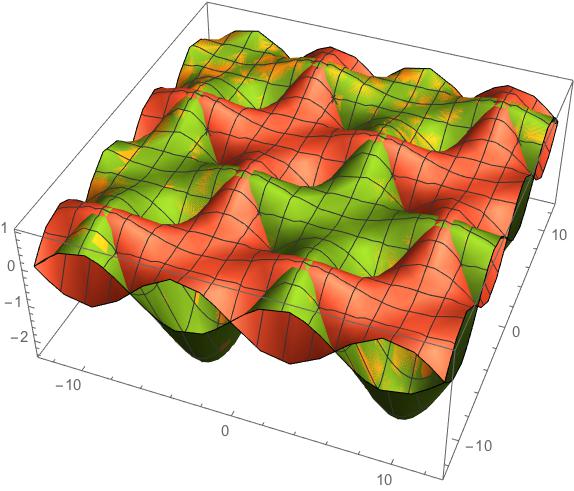}
 \includegraphics[angle=0,origin=c, scale=0.20]{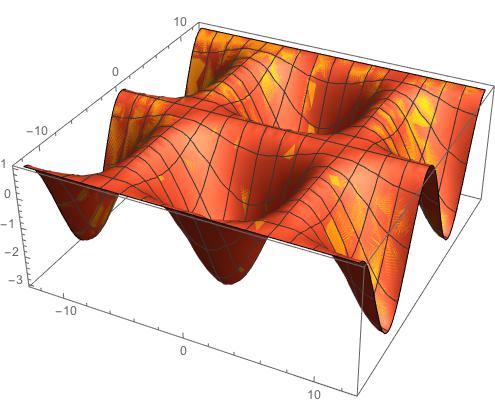}
 \label{fig:GSphasediagram}
 \caption{
 The 2d projection on $(k_x,k_y)$ plane of the spinon spectrum for the a) $U(1)$ QSL and b) QSSF states at fixed $k_z=0$.}
 \end{figure}
 
Now, we show how a Bose-Einstein condensate of bosonic spinons can give rise to a quantum spin superfluid state, a new state of matter that we propose in this work, rather than some form of magnetic order \cite{SavaryBalents}\cite{LeeOnodaBalents}\cite{GangChen}.We noted earlier that $|g_{\pm\mu}|=|\langle s^{\pm\eta_{\mathbf{r}}}_{\mathbf{r},\mathbf{r}+\eta_{\mathbf{r}}\mathbf{e}_{\mu}}\rangle|\neq 0$.The expectation value of the spin operator however vanishes because with $S^{+}_{\mathbf{r}+\mathbf{e}_{\mu}/2}\equiv  S^{x}_{\mathbf{r}+\mathbf{e}_{\mu}/2}+ i S^{y}_{\mathbf{r}+\mathbf{e}_{\mu}/2}=b^{\dag}_{\mathbf{r}}s^{+}_{\mathbf{r},\mathbf{r}+\mathbf{e}_{\mu}}b_{\mathbf{r}+\mathbf{e}_{\mu}}$ \cite{LeeOnodaBalents}, we have 
\begin{equation}
\langle S^{+}_{\mathbf{r}+\mathbf{e}_{\mu}/2}\rangle =\langle b^{\dag}_{\mathbf{r}}b_{\mathbf{r}+\mathbf{e}_{\mu}}\rangle\langle s^{+}_{\mathbf{r},\mathbf{r}+\mathbf{e}_{\mu}}\rangle= 0
\end{equation}
given that $\langle b^{\dag}_{\mathbf{r}}b_{\mathbf{r}+\mathbf{e}_{\mu}}\rangle =0$ as the immediate consequence of the translational invariance of liquid state of spins, be it spin superfluid or `normal' spin liquid.This, combined with the conditions $|\Psi|=|\langle b\rangle|=|\langle b^{\dag}\rangle|\neq 0$, $\langle s^z\rangle=0$ self-consistently concluded earlier, implies that we find a state with Bose-Einstein condensation of bosonic spinons but which does not give a magnetic order.Instead, since the state has a vanishing spin expectation value but a finite spin superfluid density, it gives a quantum spin superfluid (QSSF) state, a new state of matter that we find in this work.Having $\langle b^{\dag}_{\mathbf{r}}b_{\mathbf{r}+\eta_{\mathbf{r}}\mathbf{e}_{\mu}}\rangle=0$ representing a translational invariant state while $\langle b_{\mathbf{r}}\rangle\neq 0$ might be surprising but can be proven by considering boson correlation function \cite{SuppMat}.Furthermore, just like the QSL, the QSSF found here manifests a \textit{deconfined} state of spinons that corresponds to its propagating, fluid nature \cite{DeconfinedHiggs}.

\textit{`Bilinear' Gauge Mean-Field Theory:Spinon Dispersion, Variational Energy Calculation, Spin Superfluid Properties, and Ground State Phase Diagram.\textemdash}
To further verify the `unilinear' gauge mean-field analysis, we perform a variational energy calculation by computing the expectation value of the Hamiltonian Eq.(\ref{latticeQED}) $\langle H_{\mathrm{QED}}\rangle$ following bilinear operator mean-field decomposition \cite{LeeOnodaBalents}, but focusing only on the liquid states in which case $\langle b^{\dag}b\rangle=\langle b b\rangle=0$ \cite{SuppMat}.From the energy dispersion $\epsilon(\mathbf{k})$ of the bosonic spinon, illustrated in Fig.1, we find that the $U(1)$ QSL and the spin superfluid state are distinguished by the nature of their bosonic spinon energy spectrum $\epsilon(\mathbf{k})$; as Fig. 1 shows, while there is in general an energy gap between the `singlet' ground state and the three-fold degenerate `triplet' state (where the sublattice plays the role of the spin) in the $U(1)$ QSL (except on a surface in the 3d $\mathbf{k}$ space), the spectrum is gapless everywhere for the spin superfluid state.This extensive gaplessness drives the spinon condensation even in the absence of translational symmetry-breaking and sublattice-mixing particle-hole order.Furthermore, the intuitive prediction from the unilinear gauge mean-field analysis regarding the sign of $J_{\pm}$ and the associated flux that can give rise to spin superfluidity remarkably agrees fully with a rigorous variational energy calculation \cite{SuppMat}.Our variational energy calculation decisively shows that the quantum spin superfluid state with its $\pi$-flux ansatz for antiferromagnetic case ($J_{\pm}<0$) has lower energy than that of the $U(1)$ QSL state in an appropriate regime of the coupling $J$'s space, while this does not hold for the ferromagnetic ($J_{\pm}>0$) case \cite{SuppMat}. 

Using the energy dispersion $\epsilon(\mathbf{k})$, we can compute the first critical velocity $v_{c1}$ of the spin superfluid given by effective `phonon' velocity $\mathbf{v}_{c1}=\partial \epsilon(\mathbf{k})/\partial\mathbf{k}|_{\mathbf{k}_0}$, where $\mathbf{k}_0=2\pi(101)$ (and symmetry-related points) is the wave vector at the global minimum of $\epsilon(\mathbf{k})$ \cite{SuppMat}.This nonzero $\mathbf{k}_0$ is the origin of the deconfinement and the propagation of the spinons even when they condense as concluded earlier.The flat part at the `crossroad' in the spinon dispersion Fig.1b) serves as the `roton' that provides a second critical velocity $\mathbf{v}_{c2}=\Delta\epsilon(\mathbf{k})/\Delta\mathbf{k}$ with $\Delta\mathbf{k}=\mathbf{k}_{\mathrm{roton}}-\mathbf{k}_0$ and thence $v_{c}=\mathrm{min}[v_{c1},v_{c2}]$.From the boson phase field Eq.(\ref{phasefield}), the superfluid velocity \cite{LandauLifshitz} is given by
\begin{equation}\label{sfproperties}
\mathbf{v}_{sf}=-i\frac{\hbar}{2m_b|\Psi|^2}\left(\Psi^*\nabla\Psi-\Psi\nabla\Psi^*\right)=\frac{\hbar}{m_b}\nabla\theta=\frac{\hbar}{m_b}\mathbf{G}
\end{equation}
where $m_b$ is the effective mass of the bosonic spinon given by
\begin{equation}\label{bosonicspinonmass}
m_b=\left(\frac{\partial^2\epsilon(\mathbf{k})}{\partial\mathbf{k}^2}|_{\mathbf{k}_0}\right)^{-1}=\frac{2}{3J_{\pm}}
\end{equation}
The spin superflow is stable only when $v_{sf}<v_c$ \cite{Landau}, and this constrains the reciprocal lattice wave vector $\mathbf{G}$ in Eq.(\ref{sfproperties}); only the few smallest $\mathbf{G}$'s satisfying $0<|\mathbf{G}|<|\mathbf{G}_c|$ where
\begin{equation}\label{criticalvector}
|\mathbf{G}_c|=\frac{|\nabla_{\mathbf{k}}\epsilon(\mathbf{k})|_{\mathbf{k}_0}}{|\nabla^2_{\mathbf{k}}\epsilon(\mathbf{k})|_{\mathbf{k}_0}}
\end{equation}
which is finite for finite geometry, contribute to a stable spin superflow current \cite{SuppMat}.

\begin{figure}
 \includegraphics[angle=0,origin=c, scale=0.25]{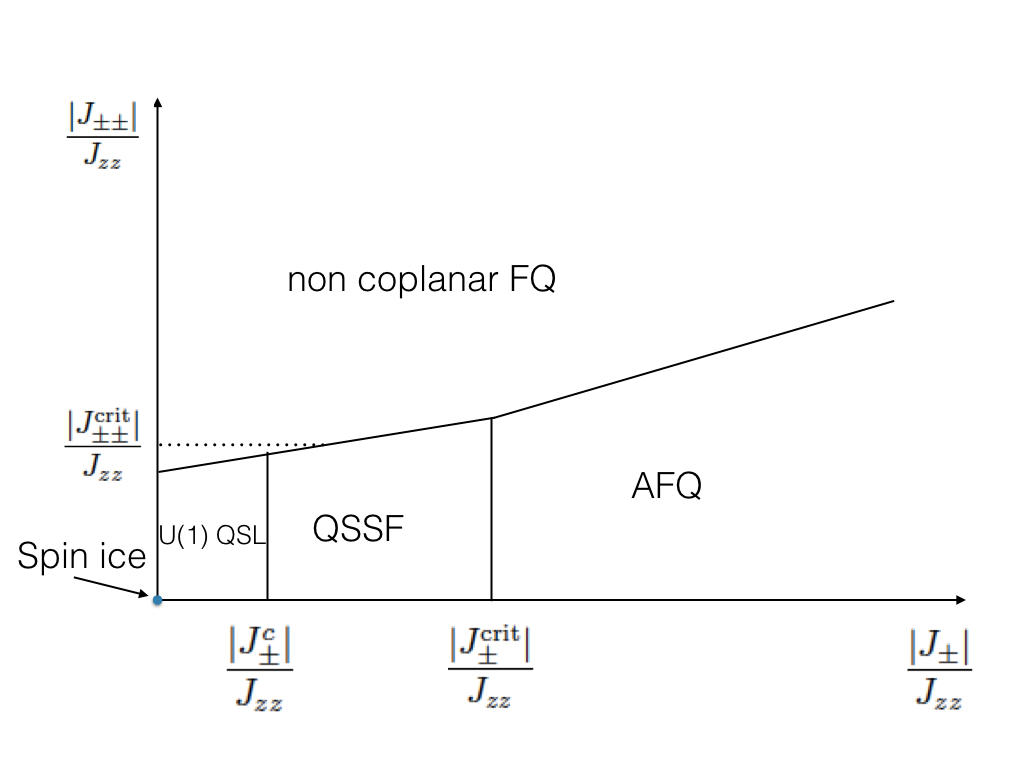}
 \label{fig:GSphasediagram}
 \caption{
 The schematic ground state phase diagram of quantum spin ice with its spin superfluid state for $J_{\pm}<0$ at $J_{z\pm}=0$.}
 \end{figure} 

It has been shown that as one increases the transverse spin coupling $J_{\pm}$ at small $J_{\pm\pm}$ starting from a $U(1)$ QSL state, bosonic spinons will condense and give rise to a translational invariance-breaking sublattice-mixing particle-hole order, which manifests as a antiferromagnetic quadrupolar (AFQ) order while increasing $J_{\pm\pm}$ to large enough value gives a noncoplanar ferroquadrupolar (FQ) order \cite{LeeOnodaBalents}.Our result indicates that a translational invariant bosonic spinon condensate state can preempt the QSL-AFQ transition and manifests as a quantum spin superfluid state.Based on variational energy calculation \cite{SuppMat} and the spin superfluid density Eq.(\ref{quantumspinsuperfluiddensity}), we conclude that the quantum spin superfluid occurs between the $U(1)$ QSL and the AFQ order while bordering the FQ order directly.We conjecture a schematic ground state phase diagram of quantum spin ice with quantum spin superfluid state in it that is illustrated in Fig.2, motivated by the result in \cite{LeeOnodaBalents}.Based on Eqs.(\ref{superfluidOP}),(\ref{SFdensity}),(\ref{sfproperties}),(\ref{bosonicspinonmass}) and our variational energy calculation \cite{SuppMat}, the QSSF should occur below a critical $J_{\pm\pm}$ line and above a critical $J_{\pm}$ value with spin superfluid density $|\Psi|^2=|J_{\pm}|/(3J_{\pm\pm})$ for $|J_{\pm}|>|J^c_{\pm}|$ \cite{regularization}.There is thus a jump in the spin superfluid density 
\begin{equation}\label{SFDjump}
\delta|\Psi|^2=\frac{|J^c_{\pm}|}{3J_{\pm\pm}}
\end{equation}
that implies a first-order quantum phase transition out of $U(1)$ QSL induced by quantum fluctuations \cite{Makhfudz}.The quantum phase transition between the QSSF and AFQ (noncoplanar FQ) occurs because as one increases $J_{\pm}(J_{\pm\pm})$ past a critical $J^{\mathrm{crit}}_{\pm}(J^{\mathrm{crit}}_{\pm\pm})$, a translational symmetry-breaking particle-hole order parameter that mixes the two sublattices of the dual diamond lattice (corresponding to the `up' and `down' tetrahedra) develops a nonzero expectation value $\langle b^{\dag}_{\mathbf{r}}b_{\mathbf{r}+\eta_{\mathbf{r}}\mathbf{e}_{\mu}}\rangle\neq 0$, which destroys the translationally invariant spin superfluid state; making it a spin analog of solid, while the Bose-Einstein condensation of bosonic spinons survives $\langle b\rangle,\langle b^{\dag}\rangle\neq 0 $ in both phases.The key distinctions between the $U(1)$ QSL, QSSF, and the AFQ/FQ phases are shown in Table 1.

\bigskip
\begin{tabular}{l*{6}{c}r}
State              & $\langle s^z_{\mathbf{r},\mathbf{r}^{\pm}_{\mu}}\rangle$ & $\langle s^{\pm}_{\mathbf{r},\mathbf{r}^{\pm}_{\mu}}\rangle$ & $\langle b_{\mathbf{r}}\rangle$ & $\langle b_{\mathbf{r}} b_{\mathbf{r}} \rangle$  & $\langle b^{\dag}_{\mathbf{r}}b_{\mathbf{r}^{\pm}_{\mu}} \rangle$  & $\langle S^z_{\tilde{\mathbf{r}}_{\mu}}\rangle$ & $\langle S^{\pm}_{\tilde{\mathbf{r}}_{\mu}} \rangle$ \\
\hline
$U(1)$ QSL & $0$ & $\neq 0$ & $0$ & $0$ & $0$ & $0$ & $0$  \\
QSSF            & $0$ & $\neq0$ & $\neq0$ & 0 &  0 & 0 &  0  \\
AFQ/FQ           & $0$ & $\neq 0$ & $\neq 0$ & $\neq 0$ & $\neq 0$ & 0 &  $\neq0$  \\
\end{tabular}
where $\mathbf{r}^{\pm}_{\mu}=\mathbf{r}\pm \mathbf{e}_{\mu}$ gives the position vector of the centers of the tetrahedra (the sites of the dual diamond lattice) while $\tilde{\mathbf{r}}_{\mu}=\mathbf{r}+\eta_{\mathbf{r}}\mathbf{e}_{\mu}/2$ gives the pyrochlore lattice site (the corner of a tetrahedron) position vector.It is to be noted that while $\langle b_{\mathbf{r}}\rangle$ is nonzero both in the QSSF and AFQ/FQ phases, this quantity is lattice-translational invariant in QSSF as the consequence of Eq.(\ref{phasefield}) but this is not the case in AFQ/FQ.

\bigskip 

\textit{Discussion.\textemdash}
\textit{From Bose-Einstein Condensation of Bosonic Spinons to Spin Superfluidity.\textemdash}
The possibility of Bose-Einstein condensation in spin systems is an actively pursued idea \cite{GiamarchiRueggTchernyshyov}.In this case, the bosons are the magnons; the quanta of spin waves around a magnetically ordered ground state excited by thermal fluctuations at low temperatures.Their condensation normally gives rise to some type of magnetic order \cite{MatsubaraMatsuda}\cite{Oshikawa} while its relation to (finite temperature) spin superfluidity is an active topic \cite{Sonin}, but spin superfluidity from magnon condensation emerging in appropriate spintronics setup has been proposed (e.g. \cite{Tserkovnyak}).In our work, we consider condensation of spinons; fractionalized excitations of quantum spin system \cite{Balents} and find that such condensation may give rise to a new state of matter; a quantum spin superfluid, rather than some type of magnetic order.

The relation between the Bose-Einstein condensation and superfluidity is itself an interesting problem \cite{LeggettRMP1999}\cite{CNYangODLRO}.In superfluid Helium, it is suggested that the superfluidity is caused by the Bose-Einstein condensation of the Helium atoms \cite{London1938}\cite{Bogoliubov1947}\cite{LeggettSFSS} but the effort in understanding this remains in progress \cite{Stringaribook}.In our work, we note that since the Bose-Einstein condensate $\langle b\rangle$ is translationally invariant and gives rise to a spin rotationally invariant state, it must correspond to a liquid state of the spins, which however cannot be a `normal' spin liquid, because it has some condensation of deconfined (propagating) spinons with finite spin superfluid velocity.We thus conclude that this state is a spin superfluid state.This state is also found to be energetically favorable over the `normal' $U(1)$ quantum spin liquid state in the fully antiferromagnetic non-Kramers quantum spin ice on an appropriate regime of the spin coupling parameter space.Other than of fundamental interest, the quantum spin superfluid state if realized experimentally could open up a prospect for dissipationless spin transport applications, even though advanced low temperature cooling technology might have to be employed.We conclude the story by reflecting that the system of charged bosons coupled to (emergent) gauge field as realized in quantum spin ice is a fascinating model and may perhaps be realized in other platforms, such as ultracold atom system where the physics discussed in this work could probably be investigated.

\textit{Note added:}
While we were finishing this manuscript, we became aware of a recent work based on quantum Monte-Carlo numerical simulation of pyrochlore spin ice at finite temperature and magnetic field along $[111]$ direction at $J_{\pm\pm}=J_{z\pm}=0$ \cite{BojosenOnoda}.In their case, they found a `monopole superfluid' state, but which manifests a \textit{confined} Higgs condensed state of spinons, corresponding to an XY ferromagnet order.Interestingly though, their monopole superfluid state in the zero field limit occurs on the same part of $J_{\pm}$ axis as ours and they also obtain a superfluid density proportional to $|J_{\perp}|\equiv |J_{\pm}|$.Our work investigates analytically the case at finite $J_{\pm\pm},J_{z\pm}$ at zero field and finds a truly `fluidic' spin superfluid state with no magnetic order, manifesting a \textit{deconfined} state of Higgs condensed spinons.The singularity in $|\Psi|^2$ at $J_{\pm\pm}=0$ noted in \cite{regularization} along the $|J^{c}_{\pm}|<|J_{\pm}|<|J^{\mathrm{crit}}_{\pm}|$ segment in Fig.2 marks a deconfinement-confinement transition of the spinons from our QSSF state to the XY ferromagnet of \cite{BojosenOnoda}.Our work is thus distinct but complements that of \cite{BojosenOnoda}.We also just learned that a quantum spin superfluid state from condensation of hard-core bosons rather than spinons was found in pyrochlore using quantum Monte-Carlo simulation a while ago \cite{Isakov} which also indicates a first order superfluid transition.These works estimate that $|J^c_{\pm}|\simeq 0.104 J_{zz}$.A spin superfluid state in 2d spin system (kagom\'{e} spin 1/2 XY model) was also deduced from quantum Monte-Carlo simulation, but requires ring-exchange interaction and has a second-order superfluid phase transition within XY universality class \cite{Melko}.

\textit{Acknowledgements.\textemdash}I.M. would like to thank Prof. I. Lukyanchuk for the endless encouragement all along.


\begin{widetext}

\centering
\textbf{Supplementary Material: Quantum Spin Superfluid from Bose-Einstein Condensation of Spinons in Pyrochlore Spin Ice}\\
\centering
Imam Makhfudz \\
\centering
Univ Lyon, ENS de Lyon, Univ Claude Bernard, \comment{CNRS, Laboratoire de Physique, F-69342} Lyon, France\\ 
\date{\today}
\bigskip

In this Supplementary Material, we first discuss some subtle points regarding the local gauge invariance in our system and the possibility of having a deconfined Higgs condensed state of bosons coupled to gauge field, then show how translational invariance can be preserved in a condensed state of spinons, and finally give some details on mean-field and variational energy calculations.

\end{widetext}

\bigskip






\section{Local Gauge Invariance in Systems with Non-Local Interactions and the Deconfined Higgs Condensate of Bosonic Spinons}
A subtle point to note regarding the effective Hamiltonian for bosonic spinon coupled to gauge field in Eq.(1) in the main text is that the $J_{\pm\pm}$ term at zeroth order level ($\langle s^+_{\mathbf{r}\mathbf{r}_{\mu}}\rangle=1+\mathcal{O}(A_{\mathbf{r}\mathbf{r}_{\mu}})$) essentially corresponds to a non-local boson-boson interaction because the boson operators are defined at three generally different sites; the centers of three adjacent tetrahedra in a pyrochlore lattice.Physically, this $J_{\pm\pm}$ term flips two spins at two corners of a tetrahedron both up ($++$) or both down ($--$) with respect to local spin vector basis defined at each corner.If we start from a state with no magnetic monopoles, this process translates into a creation of two magnetic monopoles of positive charge at the center of tetrahedron located at $\mathbf{r}$ and two negative charged monopoles at neighboring tetrahedra centers located at $\mathbf{r}+\eta_{\mathbf{r}}\mathbf{e}_{\mu}$ and $\mathbf{r}+\eta_{\mathbf{r}}\mathbf{e}_{\nu}$.This boson `scattering' process is equivalent to a nonlocal boson interaction.In the original lattice spin model, this interaction originates from the short-range nearest-neighbor spin exchange interaction, but it translates into nonlocal interaction between the bosonic spinons, despite the constant interaction coupling strength $J_{\pm\pm}$.

As a result, the Elitzur theorem \cite{Elitzur}, which permits only locally gauge invariant quantity to acquire nonzero expectation value does not apply in this system, because the theorem assumes local interaction.This is true even at gauge theory level and one does not need to resort to mean-field theory for which the local gauge invariance is broken explicitly and the Elitzur theorem ceases to apply.As a consequence, this permits a ground state with $\langle b\rangle \neq 0$ corresponding to condensation of bosonic spinons.Whether such state prevails or not can be determined from variational energy calculation.To find an instability of $U(1)$ quantum spin liquid towards a spin superfluid state (that is, the superfluid state of bosonic spinons), we perform a mean-field decomposition that explicitly breaks the $U(1)$ symmetry \cite{SachdevQPTbookS}.This gives rise to a non-conservation of local (i.e. at each site) as well as total numbers of bosons.The total \textit{charge} of bosons is however conserved because the bosons are created or annihilated in pairs of opposite charges.This is indeed the case in quantum spin ice where flipping a spin at the corner of a tetrahedron of the pyrochlore lattice creates a pair of bosonic spinons (or magnetic monopoles) of opposite signs.The spin superfluid instability thus breaks $U(1)$ boson number conservation at both local (gauge) and global levels but it conserves the total charge of the bosons; i.e., it preserves the $U(1)$ boson charge gauge invariance.The presence of nonlocal boson interaction and the preservation of the $U(1)$ boson charge gauge invariance turn out to play role in the deconfinement of the bosonic spinons, as explained below.

In the existing studies of gauge theories where a boson is coupled to gauge field, it is argued that there occur Higgs condensed state and confined state of bosons \cite{FradkinShenker}.Both of these phases are the mechanisms to remove the gauge field from low energy theory by gapping it out.It is suggested that these two phases are continuously connected \cite{FradkinShenker} when the Higgs fields (bosons) are in the fundamental representation and all the gauge invariance has been broken.This is despite the fact that they describe two different phenomena; Higgs condensation gaps out the gauge field by condensing the bosons, e.g. by Bose-Einstein condensation, whereas confined phase refers to the confinement of the bosons.Although it sounds very sensible to think that condensed bosons are also confined, the two phenomena still have distinct physical meaning; it is not impossible to have condensed bosons that are not confined, that is, the bosons are deconfined so that they can still move around.BCS theory superconductivity \cite{BCSs} gives a good example of boson condensation where the bosons (the Cooper pairs) are not confined at all, otherwise we will not get a superconducting state.There are at least two justifications for having a deconfined Higgs condensed boson state.First, if there is still unbroken gauge invariance.In our system, the $U(1)$ boson number gauge invariance is broken corresponding to the creation or annihilation of bosonic spinons but the $U(1)$ boson charge gauge invariance is still preserved, since the bosonic spinons are created or annihilated in pairs of opposite charges, so that their total charge is still conserved.Second, the presence of nonlocal, long-range interaction, as alluded to above, which was not considered in \cite{FradkinShenker}.This nonlocal interaction is produced by the $J_{\pm\pm}$ term which when it is nonzero, as we concluded in the main text, gives rise to the deconfined state of spinons corresponding to the quantum spin superfluid state that we find, but once it is zero, the resulting state is a confined `monopole superfluid' which corresponds to ordinary magnetic order (XY ferromagnet) as found in \cite{BojosenOnodaS}.

\section{Translational Invariance in a Condensed State of Spinons}
Consider boson correlation $\langle b^{\dag}_{\mathbf{r}}b_{\mathbf{r}'}\rangle$.Taking its Fourier transform, we have
\[
\langle b^{\dag}_{\mathbf{r}}b_{\mathbf{r}'}\rangle=\frac{1}{N}\sum_{\mathbf{k},\mathbf{k}'}\langle b^{\dag}_{\mathbf{k}}b_{\mathbf{k}'}\rangle e^{i(\mathbf{k}'\cdot\mathbf{r}'-\mathbf{k}\cdot\mathbf{r})}
\]
\begin{equation}
=\frac{1}{N}\sum_{\mathbf{k},\mathbf{q}}\langle b^{\dag}_{\mathbf{k}}b_{\mathbf{k}+\mathbf{q}}\rangle e^{i\mathbf{q}\cdot\mathbf{r}}e^{i(\mathbf{k}+\mathbf{q})\cdot(\mathbf{r}'-\mathbf{r})}
\end{equation}
Now, consider the condensed boson state with $\langle b_{\mathbf{r}}\rangle=|\Psi|\exp(i\mathbf{G}\cdot\mathbf{r})$ that we have found.This $\langle b_{\mathbf{r}}\rangle$ is (lattice) translational invariant (in the dual diamond lattice).A translational invariant state, by definition, has no density wave order in terms of the above boson correlation.That is $\langle b^{\dag}_{\mathbf{k}}b_{\mathbf{k}+\mathbf{q}}\rangle=\langle b^{\dag}_{\mathbf{k}}b_{\mathbf{k}}\rangle \delta(\mathbf{q})$ so that
\begin{equation}
\langle b^{\dag}_{\mathbf{r}}b_{\mathbf{r}'}\rangle=\frac{1}{N}\sum_{\mathbf{k}}\langle b^{\dag}_{\mathbf{k}}b_{\mathbf{k}}\rangle e^{i\mathbf{k}\cdot(\mathbf{r}'-\mathbf{r})}
\end{equation}

Suppose that at the ground state ($T=0$) we can have a total condensation (this is indeed the case in ideal Bose gas) and that the boson dispersion has periodic lattice of global minima located at $\mathbf{k}_0$ and the symmetry-related points (as is the case in our work, c.f. Fig. 1 in the main text), such that all the bosons are condensed at those wave vectors.Then, the boson correlation function becomes
\begin{equation}
\langle b^{\dag}_{\mathbf{r}}b_{\mathbf{r}'}\rangle=\frac{1}{N}\sum_{\mathbf{k}_0's}\langle b^{\dag}_{\mathbf{k}_0}b_{\mathbf{k}_0}\rangle e^{i\mathbf{k}_0\cdot(\mathbf{r}'-\mathbf{r})}
\end{equation}
where $\mathbf{k}_0's$ denotes $\mathbf{k}_0$ and all the wave vectors symmetry related to $\mathbf{k}_0$.It is also intuitively clear to expect that $\langle b^{\dag}_{\mathbf{k}}b_{\mathbf{k}}\rangle$ is the same for all those wave vectors, since their dispersions are precisely identical to each other.Now, if $\mathbf{k}_0=0$, using the standard random phase assumption, it is clear that the contributions from all other symmetry-related points cancel out each other but we are still left with the `zero mode' contribution from $\mathbf{k}_0$ that results in $\langle b^{\dag}_{\mathbf{r}}b_{\mathbf{r}'}\rangle\neq 0$.However, if $\mathbf{k}_0\neq 0$ as is the case with our QSSF state ($\mathbf{k}_0=2\pi(101)$), there is no `zero mode' contribution and all the terms in the sum cancel out each other to zero; $\langle b^{\dag}_{\mathbf{r}}b_{\mathbf{r}'}\rangle=0$.This can also be verified explicitly by substituting $\mathbf{k}_0=2\pi(101)$ and $\mathbf{r}'-\mathbf{r}=\mathbf{e}_{\mu}$ which gives $\exp(i\mathbf{k}_0\cdot(\mathbf{r}'-\mathbf{r}))=\pm 1$ and yields exact cancellation to zero upon summation.This proves our assertion that we can have translational invariant state characterized by $\langle b^{\dag}_{\mathbf{r}}b_{\mathbf{r}+\eta_{\mathbf{r}}\mathbf{e}_{\mu}}\rangle=0$ even with $\langle b_{\mathbf{r}}\rangle\neq 0$.This conclusion remains valid even if we do not have total condensation, as long as the condensed bosons are still distributed uniformly among the global minima $\mathbf{k}_0's$ (which is a valid assumption in the absence of time reversal or parity breaking), we still have $\langle b^{\dag}_{\mathbf{r}}b_{\mathbf{r}'}\rangle= 0$ for $\mathbf{k}_0\neq 0$.The sum of the contributions from the uncondensed bosons vanishes due to the random phase assumption.

\section{Gauge Mean-Field Theory Equations}
\subsection{Mean-Field Decoupling}
From Eq.(1) in the main text, we perform the following `unilinear' mean-field decomposition that stresses the $U(1)$-breaking character of a superfluid state \cite{SachdevQPTbookS}: 
\begin{equation}
s^z_{\mathbf{r}\mathbf{r}_{\mu}}s^z_{\mathbf{r}\mathbf{r}_{\nu}}\rightarrow s^z_{\mathbf{r}\mathbf{r}_{\mu}}\langle s^z_{\mathbf{r}\mathbf{r}_{\nu}}\rangle +\langle s^z_{\mathbf{r}\mathbf{r}_{\mu}}\rangle s^z_{\mathbf{r}\mathbf{r}_{\nu}}-\langle s^z_{\mathbf{r}\mathbf{r}_{\mu}}\rangle\langle s^z_{\mathbf{r}\mathbf{r}_{\nu}}\rangle
\end{equation}
\[
s^+_{\mathbf{r}\mathbf{r}_{\nu}}s^-_{\mathbf{r}\mathbf{r}_{\mu}}b^{\dag}_{\mathbf{r}_{\mu}}b_{\mathbf{r}_{\nu}}\rightarrow \langle s^+_{\mathbf{r}\mathbf{r}_{\nu}}\rangle \langle s^-_{\mathbf{r}\mathbf{r}_{\mu}}\rangle (b^{\dag}_{\mathbf{r}_{\mu}}\langle b_{\mathbf{r}_{\nu}}\rangle + \langle b^{\dag}_{\mathbf{r}_{\mu}}\rangle b_{\mathbf{r}_{\nu}}   - \langle b^{\dag}_{\mathbf{r}_{\mu}} \rangle\langle b_{\mathbf{r}_{\nu}} \rangle)
\]
\begin{equation}
+(s^+_{\mathbf{r}\mathbf{r}_{\nu}}\langle s^-_{\mathbf{r}\mathbf{r}_{\mu}} \rangle + \langle s^+_{\mathbf{r}\mathbf{r}_{\nu}}\rangle s^-_{\mathbf{r}\mathbf{r}_{\mu}}-\langle s^+_{\mathbf{r}\mathbf{r}_{\nu}}\rangle\langle s^-_{\mathbf{r}\mathbf{r}_{\mu}}\rangle)\langle b^{\dag}_{\mathbf{r}_{\mu}}\rangle\langle b_{\mathbf{r}_{\nu}}\rangle
\end{equation} 
\[
s^+_{\mathbf{r}\mathbf{r}_{\mu}}s^+_{\mathbf{r}\mathbf{r}_{\nu}}b^{\dag}_{\mathbf{r}}b^{\dag}_{\mathbf{r}}b_{\mathbf{r}_{\mu}}b_{\mathbf{r}_{\nu}}\rightarrow 
\]
\[
\langle s^+_{\mathbf{r}\mathbf{r}_{\mu}}\rangle\langle s^+_{\mathbf{r}\mathbf{r}_{\nu}}\rangle\left[ 6b^{\dag}_{\mathbf{r}}\langle b^{\dag}_{\mathbf{r}}\rangle\langle b_{\mathbf{r}_{\mu}}\rangle\langle b_{\mathbf{r}_{\nu}}\rangle+3\langle b^{\dag}_{\mathbf{r}}\rangle^2(b_{\mathbf{r}_{\mu}}\langle b_{\mathbf{r}_{\nu}}\rangle+b_{\mathbf{r}_{\nu}}\langle b_{\mathbf{r}_{\mu}}\rangle)\right]
\]
\[
+3\langle b^{\dag}_{\mathbf{r}}\rangle^2\langle b_{\mathbf{r}_{\mu}}\rangle\langle b_{\mathbf{r}_{\nu}}\rangle(s^+_{\mathbf{r}\mathbf{r}_{\mu}}\langle s^+_{\mathbf{r}\mathbf{r}_{\nu}}\rangle+s^+_{\mathbf{r}\mathbf{r}_{\nu}}\langle s^+_{\mathbf{r}\mathbf{r}_{\mu}}\rangle)
\]
\begin{equation}
-9\langle b^{\dag}_{\mathbf{r}}\rangle^2\langle b_{\mathbf{r}_{\mu}}\rangle\langle b_{\mathbf{r}_{\nu}}\rangle\langle s^+_{\mathbf{r}\mathbf{r}_{\mu}}\rangle\langle s^+_{\mathbf{r}\mathbf{r}_{\nu}}\rangle
\end{equation}
\[
S^z_{\mathbf{r}\mathbf{r}_{\mu}}s^{\pm}_{\mathbf{r}\mathbf{r}_{\nu}}b^{\dag}_{\mathbf{r}}b_{\mathbf{r}_{\nu}}\rightarrow \langle S^z_{\mathbf{r}\mathbf{r}_{\mu}}\rangle \langle s^{\pm}_{\mathbf{r}\mathbf{r}_{\nu}}\rangle (b^{\dag}_{\mathbf{r}}\langle b_{\mathbf{r}_{\nu}}\rangle+\langle b^{\dag}_{\mathbf{r}}\rangle b_{\mathbf{r}_{\nu}})
\]
\[
+\langle b^{\dag}_{\mathbf{r}}\rangle\langle b_{\mathbf{r}_{\nu}}\rangle(S^z_{\mathbf{r}\mathbf{r}_{\mu}}\langle s^{\pm}_{\mathbf{r}\mathbf{r}_{\nu}}\rangle+\langle S^z_{\mathbf{r}\mathbf{r}_{\mu}}\rangle s^{\pm}_{\mathbf{r}\mathbf{r}_{\nu}})
\]
\begin{equation}
-2\langle S^z_{\mathbf{r}\mathbf{r}_{\mu}}\rangle\langle s^{\pm}_{\mathbf{r}\mathbf{r}_{\nu}}\rangle\langle b^{\dag}_{\mathbf{r}}\rangle\langle b_{\mathbf{r}_{\nu}}\rangle
\end{equation}
where we have single boson creation or annihilation operators rather than bilinear products of them.It is also to be noted that $\langle b^{\dag}\rangle,\langle b\rangle \neq 0$ in general because the nonlocal boson $J_{\pm\pm}$ interaction annuls the applicability of Elitzur theorem.If we further assume that the boson expectation value does not break translational symmetry, we can simplify the mean-field decomposition further.Denoting $\langle b\rangle = \Psi,\langle b^{\dag}\rangle = \Psi^{\dag}$, we get
\[
s^+_{\mathbf{r}\mathbf{r}_{\nu}}s^-_{\mathbf{r}\mathbf{r}_{\mu}}b^{\dag}_{\mathbf{r}_{\mu}}b_{\mathbf{r}_{\nu}}\rightarrow \langle s^+_{\mathbf{r}\mathbf{r}_{\nu}}\rangle \langle s^-_{\mathbf{r}\mathbf{r}_{\mu}}\rangle (b^{\dag}_{\mathbf{r}_{\mu}}\Psi + \Psi^* b_{\mathbf{r}_{\nu}}   - |\Psi|^2)
\]
\begin{equation}
+(s^+_{\mathbf{r}\mathbf{r}_{\nu}}\langle s^-_{\mathbf{r}\mathbf{r}_{\mu}} \rangle + \langle s^+_{\mathbf{r}\mathbf{r}_{\nu}}\rangle s^-_{\mathbf{r}\mathbf{r}_{\mu}}-\langle s^+_{\mathbf{r}\mathbf{r}_{\nu}}\rangle\langle s^-_{\mathbf{r}\mathbf{r}_{\mu}}\rangle)|\Psi|^2
\end{equation} 
\[
s^+_{\mathbf{r}\mathbf{r}_{\mu}}s^+_{\mathbf{r}\mathbf{r}_{\nu}}b^{\dag}_{\mathbf{r}}b^{\dag}_{\mathbf{r}}b_{\mathbf{r}_{\mu}}b_{\mathbf{r}_{\nu}}\rightarrow 
\]
\[
\langle s^+_{\mathbf{r}\mathbf{r}_{\mu}}\rangle\langle s^+_{\mathbf{r}\mathbf{r}_{\nu}}\rangle\left[ 6b^{\dag}_{\mathbf{r}}|\Psi|^2\Psi+3|\Psi|^2\Psi^*(b_{\mathbf{r}_{\mu}}+b_{\mathbf{r}_{\nu}})\right]
\]
\begin{equation}
+3|\Psi|^4(s^+_{\mathbf{r}\mathbf{r}_{\mu}}\langle s^+_{\mathbf{r}\mathbf{r}_{\nu}}\rangle+s^+_{\mathbf{r}\mathbf{r}_{\nu}}\langle s^+_{\mathbf{r}\mathbf{r}_{\mu}}\rangle)
-9|\Psi|^4\langle s^+_{\mathbf{r}\mathbf{r}_{\mu}}\rangle\langle s^+_{\mathbf{r}\mathbf{r}_{\nu}}\rangle
\end{equation}
\[
S^z_{\mathbf{r}\mathbf{r}_{\mu}}s^{\pm}_{\mathbf{r}\mathbf{r}_{\nu}}b^{\dag}_{\mathbf{r}}b_{\mathbf{r}_{\nu}}\rightarrow \langle S^z_{\mathbf{r}\mathbf{r}_{\mu}}\rangle \langle s^{\pm}_{\mathbf{r}\mathbf{r}_{\nu}}\rangle (b^{\dag}_{\mathbf{r}}\Psi+\Psi^* b_{\mathbf{r}_{\nu}})
\]
\begin{equation}
+|\Psi|^2(S^z_{\mathbf{r}\mathbf{r}_{\mu}}\langle s^{\pm}_{\mathbf{r}\mathbf{r}_{\nu}}\rangle+\langle S^z_{\mathbf{r}\mathbf{r}_{\mu}}\rangle s^{\pm}_{\mathbf{r}\mathbf{r}_{\nu}})
-2\langle S^z_{\mathbf{r}\mathbf{r}_{\mu}}\rangle\langle s^{\pm}_{\mathbf{r}\mathbf{r}_{\nu}}\rangle |\Psi|^2
\end{equation}
Applying the above decoupling to Eq.(1) in the main text, we obtain Eqs.(2-4).

\subsection{Gauge Mean-Field Equation in the Matter Sector: Derivation of the Spin Superfluid Density}

From the gauge mean-field decoupled Hamiltonian in Eqs.(2-4) in the main text, we take $\delta \langle H^{\mathrm{MF}}_{\mathrm{QED}}(b)\rangle/\delta \langle b\rangle$. We obtain
\[
\frac{\delta \langle H^{\mathrm{MF}}_{\mathrm{QED}}(b)\rangle}{\delta \langle b_{\mathbf{r}_{\nu}}\rangle}=\sum_{\mathbf{r},\mu\neq\nu}\{-J_{\pm}g_{-\mu}g_{+\nu}\Psi^*+3J_{\pm\pm}|\Psi|^2\times
\]
\begin{equation}
\mathrm{Re}\left[\gamma^{-2\eta_{\mathbf{r}}}_{\mu\nu}g_{+\mu}g_{+\nu}\Psi^*\right]
-2J_{z\pm}\mathrm{Re}\left[\gamma^{-\eta_{\mathbf{r}}}_{\mu\nu} g_{z\mu} g_{+\nu}\Psi^*\right]\}
\end{equation}
where $\Psi=\langle b_{\mathbf{r}}\rangle$, $g_{\pm\mu}=\langle s^{\pm\eta_{\mathbf{r}}}_{\mathbf{r}\mathbf{r}_{\mu}}\rangle=\Delta \exp(\pm i \overline{A}_{\mathbf{r},\mathbf{r}_{\mu}})$ and $g_{z\mu}=\langle s^{z}_{\mathbf{r}\mathbf{r}_{\mu}}\rangle$.We then use the translationally invariant bosonic spinon condensation density ansatz $\Psi=|\Psi|\exp(i\theta(\mathbf{r}))$ and solve the equation $\delta \langle H^{\mathrm{MF}}_{\mathrm{QED}}(b)\rangle/\delta \langle b\rangle=0$ for $|\Psi|^2$, which gives Eq.(5) in the main text.

\subsection{Gauge Mean-Field Equation in the Gauge Sector}

From the gauge mean-field decoupled Hamiltonian in Eqs.(2-4) in the main text, we take $\delta \langle H^{\mathrm{MF}}_{\mathrm{QED}}(s^{\pm},s^z)\rangle /\delta \langle s^{\pm}\rangle $. We obtain
\[
\frac{\delta \langle H^{\mathrm{MF}}_{\mathrm{QED}}(s^{\pm},s^z)\rangle}{\delta \langle s^{\eta_{\mathbf{r}}}_{\mathbf{r}\mathbf{r}_{\nu}}\rangle}=-J_{\pm}\sum_{\mathbf{r},\mu\neq\nu}|\Psi|^2 g_{-\mu}+\sum_{\mathbf{r},\mu\neq\nu}|\Psi|^2\times
\]
\begin{equation}\label{dHpdspm}
\left[3J_{\pm\pm}\mathrm{Re}\left[\gamma^{-2\eta_{\mathbf{r}}}_{\mu\nu}|\Psi|^2g_{+\mu}\right]-2J_{z\pm}\mathrm{Re}\left[\gamma^{-\eta_{\mathbf{r}}}_{\mu\nu}g_{z\mu}\right]\right]
\end{equation}
Setting $\delta \langle H^{\mathrm{MF}}_{\mathrm{QED}}(s^{\pm},s^z)\rangle /\delta \langle s^{\pm}\rangle =0$, the solution for $g_{\pm\mu}=\langle s^{\pm}_{\mathbf{r}\mathbf{r}_{\mu}}\rangle$ is complex-valued in general, as stated in the main text.Indeed, this solution is complex-valued in any liquid state of spins; in both the $U(1)$ quantum spin liquid and quantum spin superfluid discussed in this work.

From the gauge mean-field decoupled Hamiltonian in Eqs.(2-4) in the main text, we take $\delta \langle H^{\mathrm{MF}}_{\mathrm{QED}}(s^{\pm},s^z)\rangle/\delta \langle s^z\rangle$. We obtain
\[
\frac{\delta \langle H^{\mathrm{MF}}_{\mathrm{QED}}(s^{\pm},s^z)\rangle}{\delta \langle s^z_{\mathbf{r}\mathbf{r}_{\mu}}\rangle}=
\]
\begin{equation}\label{dHpdsz}
J_{zz}\sum_{\mathbf{r},\nu}g_{z\nu}
-2J_{z\pm}|\Psi|^2\sum_{\mathbf{r},\nu\neq\mu}\mathrm{Re}\left[\gamma^{-\eta_{\mathbf{r}}}_{\mu\nu} g_{+\nu}\right]
\end{equation}
We then substitute our gauge mean-field ansatz given in Eq.(7) in the main text.We find that, remarkably, for the ansatz Eq.(7) in the main text, the sum $\sum_{\mu\neq\nu}$ in the second term in Eq.(\ref{dHpdsz}) above (that multiplies $J_{z\pm}$) vanishes identically.As a result, the extremum condition $\delta \langle H^{\mathrm{MF}}_{\mathrm{QED}}(s^{\pm},s^z)\rangle/\delta \langle s^z\rangle=0$ implies that $g_{z\mu}=\langle s^{z}_{\mathbf{r}\mathbf{r}_{\mu}}\rangle=0$ (if we impose uniform $g_{z\nu}$ to minimize the Ising energy term $\sim g_{z\nu}^2$), regardless of $J_{z\pm}$.As far as perturbative regime is concerned, no Ising order (where $\langle s^z\rangle\neq 0$) has been found in the existing works that employ the generic model \cite{SavaryBalentsS}\cite{LeeOnodaBalentsS}, thus corroborating the correctness of our mean-field analysis result above.Such Ising order is argued to occur in simpler XXZ model where only $J_{zz}$ and $J_{\pm}$ are taken into account \cite{GangChenS}.The logical explanation for this is that any finite amount of coupling $J_{z\pm}$ readily drives strong longitudinal spin quantum fluctuations that enforce $\langle s^{z}_{\mathbf{r}\mathbf{r}_{\mu}}\rangle=0$ while $J_{\pm\pm}$ reduces the transverse spin fluctuations and makes $\langle s^{\pm}_{\mathbf{r}\mathbf{r}_{\mu}}\rangle\neq 0$.In \cite{LeeOnodaBalentsS} it was found that the $\langle s^z\rangle=0$ state indeed minimizes the variational energy $\langle H_{\mathrm{QED}}\rangle$ for $J_{z\pm}=0$ that justifies them to neglect $J_{z\pm}$ altogether.We work under the assumption that this conclusion holds even to finite but small $J_{z\pm}/J_{zz}\ll 1$ where we are still in the translationally invariant liquid state of spins, be it QSL or QSSF, which then protects the conclusions derived from Eqs.(5-6) in the main text.This assumption is further supported by the observation that at finite but small $J_{z\pm}/J_{zz}$, one still has $U(1)$ QSL at sufficiently large $|J_{\pm}|/J_{zz}$ \cite{SavaryBalentsS}.

\section{Deduction of the Gauge Mean-Field Link Variable Ansatz and Superfluid Properties}

The phase field (average gauge field) ansatz given in Eq.(7) in the main text is derived simply by requiring that the terms under the sums $\sum_{\mathbf{r},\mu\neq \nu}$ on right hand side of Eqs.(5-6) is a real quantity (whose precise sign depends on whether $J_{\pm}>0$ (ferromagnetic case) or $J_{\pm}<0$ (antiferromagnetic case) but overall must be positive).

From Eq.(6) in the main text, we write $g_{\pm\mu}=\Delta\exp(\pm i\overline{A}_{\mathbf{r},\mathbf{r}_{\mu}})$ and $\gamma_{\mu\nu}=|\gamma_{\mu\nu}|\exp(i\phi_{\mu\nu})$.Eq.(6) in the main text then becomes
\[
|\Psi|^2=
\]
\begin{equation}\label{quantumspinsuperfluiddensityS}
\frac{J_{\pm}\sum_{\mathbf{r},\mu\neq\nu}e^{i(\overline{A}_{\mathbf{r},\mathbf{r}^{-}_{\mu}}+\overline{A}_{\mathbf{r},\mathbf{r}^{+}_{\nu}}-\theta(\mathbf{r}_{\mu}))}}{3J_{\pm\pm}\sum_{\mathbf{r},\mu\neq\nu}|\gamma_{\mu\nu}|^{-2\eta_{\mathbf{r}}}\cos(\overline{A}_{\mathbf{r},\mathbf{r}^{+}_{\mu}}+\overline{A}_{\mathbf{r},\mathbf{r}^{+}_{\nu}}-\theta(\mathbf{r}_{\mu})-2\phi_{\mu\nu}\eta_{\mathbf{r}})}
\end{equation}
where we have used $\Delta=1/2$ and $\mathbf{r}^{\pm}_{\mu}=\mathbf{r}\pm\eta_{\mathbf{r}}\mathbf{e}_{\mu}$.We have in total five unknown variables; four average gauge field $\overline{A}_{\mathbf{r}\mathbf{r}_{\mu}}$ with $\overline{A}_{\mathbf{r}\mathbf{r}-\eta_{\mathbf{r}}\mathbf{e}_{\mu}}=-\overline{A}_{\mathbf{r}\mathbf{r}+\eta_{\mathbf{r}}\mathbf{e}_{\mu}}$ and $\mu=0,1,2,3$ plus one boson phase field $\theta(\mathbf{r}_{\mu})$.The first condition to solve for these five unknown variables is obtained by requiring each term under the sum $\sum_{\mu\neq\nu}$ in the numerator of Eq.(\ref{quantumspinsuperfluiddensityS}) to be real, we have
\begin{equation}\label{numerator}
\overline{A}_{\mathbf{r},\mathbf{r}^{+}_{\nu}}-\overline{A}_{\mathbf{r},\mathbf{r}^{+}_{\mu}}-\theta(\mathbf{r}_{\mu})=\tilde{n}\pi
\end{equation}
where $\tilde{n}=0,\pm 1,\pm 2,\pm 3,\cdots$ and we have used $\overline{A}_{\mathbf{r},\mathbf{r}^{-}_{\mu}}=-\overline{A}_{\mathbf{r},\mathbf{r}^{+}_{\mu}}$.It turns out that the above equation is also the condition for the $\delta \langle H^{\mathrm{MF}}_{\mathrm{QED}}(s^{\pm},s^z)\rangle /\delta \langle s^{\pm}\rangle =0$ to be satisfied and is therefore mandatory.The second condition is fixing the boson phase field to be 
\[
\theta(\mathbf{r}_{\mu})=\mathbf{G}\cdot\mathbf{r}_{\mu}
\]
where $\mathbf{G}$ is the reciprocal lattice wave vector of the dual diamond lattice.This gives $\mathbf{G}\cdot\mathbf{r}_{\mu}=2m\pi$ with $m=0,\pm 1, \pm 2, \pm 3,\cdots$ and 
\begin{equation}\label{numerator}
\overline{A}_{\mathbf{r},\mathbf{r}^{+}_{\nu}}-\overline{A}_{\mathbf{r},\mathbf{r}^{+}_{\mu}}=n\pi
\end{equation}
where $n=\tilde{n}+m=0,\pm 1, \pm 2, \pm 3, \cdots$.A simple but yet nontrivial ansatz that satisfies Eq.(\ref{numerator}) is given in Eq.(7) in the main text.Substituting the ansatz into Eq.(\ref{quantumspinsuperfluiddensityS}), we find that the sum in the numerator is positive whereas that in the denominator is negative, which means only the $J_{\pm}<0$ (antiferromagnetic case) can give $|\Psi|^2>0$, that is, a spin superfluid state.The superfluid velocity is given by
\begin{equation}
\mathbf{v}_{sf}=-i\frac{\hbar}{2m_b|\Psi|^2}\left(\Psi^*\nabla\Psi-\Psi\nabla\Psi^*\right)=\frac{\hbar}{m_b}\nabla\theta=\frac{\hbar}{m_b}\mathbf{G}
\end{equation}
where $m_b$ is the effective mass of the bosonic spinon.This is obtained by expanding the energy dispersion of the bosonic spinon, which is computed in the following section, to quadratic order
\begin{equation}
\epsilon(\mathbf{k})=\epsilon(\mathbf{k}_0)+\Delta\mathbf{k}\cdot\nabla\epsilon(\mathbf{k})|_{\mathbf{k}_0}+\frac{1}{2}\Delta k_i\Delta k_j\frac{\partial^2\epsilon(\mathbf{k})}{\partial k_i\partial k_j}|_{\mathbf{k}_0}
\end{equation}
which gives $m_b=(\partial^2\epsilon(\mathbf{k})/\partial\mathbf{k}^2|_{\mathbf{k}_0})^{-1}$ where $\mathbf{k}_0$ corresponds to the wave vector at the global minima of the spinon energy dispersion $\epsilon(\mathbf{k})$.The linear derivative term gives the first critical velocity of `phonon' type $\mathbf{v}_{c1}=\partial \epsilon(\mathbf{k})/\partial\mathbf{k}|_{\mathbf{k}_0}$.The second critical velocity is determined by the `roton' type of excitations; local but not global minima, represented by the `flat' part in the spinon energy dispersion of the QSSF state Fig. 1b) in the main text; $\mathbf{v}_{c2}=\left(\epsilon(\mathbf{k}_{\mathrm{roton}})-\epsilon(\mathbf{k}_0)\right)/(\mathbf{k}_{\mathrm{roton}}-\mathbf{k}_0)$.The spin superfluid superflow is stable only when $v_{sf}<v_c$.The actual critical velocity is the minimum of the $v_{c1}$ and $v_{c2}$.Let us for illustration assume that $v_{c1}<v_{c2}$ so that $v_{c}=v_{c1}$, as is the usual case in superfluid Helium.The condition $v_{sf}<v_{c}$ turns out to constrain the reciprocal lattice wave vector $\mathbf{G}$; only the few smallest $|\mathbf{G}|$'s contribute to a stable spin superflow.To derive the bound, we note that while the explicit expressions for $v_{sf}$ and $v_c$ are delicate, they can be written as $v_{sf}=\alpha|J_{\pm}||\mathbf{G}|$ (we set $\hbar=1$ for compactness) and $v_c=\beta|J_{\pm}|$, where $\alpha=|\nabla^2_{\mathbf{k}}\epsilon(\mathbf{k})|_{\mathbf{k}_0}/|J_{\pm}|$ while $\beta=|\nabla_{\mathbf{k}}\epsilon(\mathbf{k})|_{\mathbf{k}_0}/|J_{\pm}|$.Substituting these into $v_{sf}<v_c$, we obtain $|\mathbf{G}|<|\mathbf{G}_c|$ where $|\mathbf{G}_c|=\beta/\alpha=|\nabla_{\mathbf{k}}\epsilon(\mathbf{k})|/|\nabla^2_{\mathbf{k}}\epsilon(\mathbf{k})|_{\mathbf{k}_0}$.The minima of $\epsilon(\mathbf{k})$ occur at $\mathbf{k}_0=2\pi(101)$ modulo the reciprocal lattice vector $\mathbf{G}$, which gives $m_b=2/(3J_{\pm})$ corresponding to $\alpha=3/2$.The critical velocity $v_{c1}$ however is found to be zero, which originates from a cosine profile of the energy dispersion around the minima, so that the linear component is zero.This latter point is valid only for infinite system; for finite system, e.g. in a thick slab or thin film geometry normal to the $\langle 111\rangle$ direction for example, the momentum space will have an infrared cutoff $\delta \mathbf{k}$ and the critical velocity is not defined at $\mathbf{k}_0$, but at $\mathbf{k}_0\pm \delta\mathbf{k}$ and this will give a nonzero critical velocity $v_{c1}$.This is really reminiscent of the Berezinskii-Kosterlitz-Thouless theory that explains the occurrence of superfluidity in thin film of liquid helium despite the absence of long-range order \cite{BKT}. 

\section{Variational Energy Calculation}
The most radical feature of the quantum spin superfluid, if one wishes, is that it breaks neither spin rotational symmetry nor translational symmetry, but yet it breaks gauge invariance associated with boson number conservation.In the context of gMFT \cite{SavaryBalentsS}\cite{LeeOnodaBalentsS}, this implies that $\langle b\rangle \neq 0$.We will verify whether such state is energetically favorable, using the complex average gauge field ansatz found in Eq.(7) in the main text.Since as discussed in the main text the supposed spin superfluid should, if ever exists, appear next to the spin liquid, the translational symmetry-breaking particle-hole order parameter $\langle b^{\dag}_{\mathbf{r}}b_{\mathbf{r}\pm\mathbf{e}_{\mu}}\rangle$ is supposed to vanish.In this case, the only order parameter of interest is the link gauge field expectation value $\langle s^{\pm}\rangle$, which is given by Eq.(7) in the main text.The other order parameters $\langle b_{\mathbf{r}} b_{\mathbf{r}} \rangle$ and $\langle b^{\dag}_{\mathbf{r}}b_{\mathbf{r}^{\pm}_{\mu}} \rangle$ appearing in the quadratic operator mean-field Hamiltonian \cite{LeeOnodaBalentsS} are zero as well for the $U(1)$ QSL and our proposed quantum spin superfluid (QSSF) state.

We start from the Hamiltonian Eq.(1) in the main text and perform mean-field decoupling that produces bilinear operator Hamiltonian \cite{LeeOnodaBalentsS}, which we referred to as `bilinear gauge mean-field theory' in the main text.We focus on the bosonic spinon mean-field Hamiltonian because it is the bosonic spinon (variational) energy that we want to minimize.The Hamiltonian is given by
\[
H_b=\frac{J_{zz}}{2}\sum_{\mathbf{r}}Q^2_{\mathbf{r}}-\frac{J_{\pm}}{4}\sum_{\mathbf{r},\mu\neq\nu}b^{\dag}_{\mathbf{r}_{\mu}}b_{\mathbf{r}_{\nu}}e^{i(\overline{A}_{\mathbf{r}\mathbf{r}^+_{\nu}}-\overline{A}_{\mathbf{r},\mathbf{r}^{+}_{\mu}})}
\]
The $J_{\pm\pm}$ part drops out because the corresponding particle-hole order parameter of the spinons that mix the two sublattices and pairing order parameters of the spinons on the same sublattice vanish for the $U(1)$ QSL and our QSSF states of our interest.The $J_{z\pm}$ part also drops out because we have $\langle s^z\rangle = 0$.We then write the action given by
\begin{widetext}
\[
S[Q,b]=\int d\tau\left(\sum_{\mathbf{r}}Q_{\mathbf{r}}\partial_{\tau}b_{\mathbf{r}}-H_b\right)=\int d\tau\left[\sum_{\mathbf{r}}Q_{\mathbf{r}}\partial_{\tau}b_{\mathbf{r}}-\frac{J_{zz}}{2}\sum_{\mathbf{r}}Q^2_{\mathbf{r}}+\frac{J_{\pm}}{4}\sum_{\mathbf{r},\mu\neq\nu}b^{\dag}_{\mathbf{r}_{\mu}}b_{\mathbf{r}_{\nu}}e^{i(\overline{A}_{\mathbf{r}\mathbf{r}^+_{\nu}}-\overline{A}_{\mathbf{r},\mathbf{r}^{+}_{\mu}})}+\sum_{\mathbf{r}}\lambda_{\mathbf{r}}(|b_{\mathbf{r}}|^2-1)\right]
\]
\end{widetext}
where the Lagrange multiplier $\lambda$ term is introduced explicitly to impose the constraint $|b_{\mathbf{r}}|^2=1$.We then integrate out the $Q_{\mathbf{r}}$ in the partition function $Z=\int DQ\int Db\int D\lambda \exp(S[Q,b,\lambda])=\int Db \int D\lambda \exp(S[b,\lambda])$.The result is the effective action for the spinons
\[
S[b,\lambda]=\int d\tau\left[\frac{1}{2J_{zz}}\sum_{\mathbf{r}}|\partial_{\tau}b_{\mathbf{r}}|^2-\frac{J_{\pm}}{4}\sum_{\mathbf{r},\mu\neq\nu}b^{\dag}_{\mathbf{r}_{\mu}}b_{\mathbf{r}_{\nu}}e^{i\Delta\overline{A}_{\mu\nu}}\right]
\]
\begin{equation}
+\int d\tau\sum_{\mathbf{r}}\lambda_{\mathbf{r}}(|b_{\mathbf{r}}|^2-1)
\end{equation}
where we have defined $\Delta\overline{A}_{\mu\nu}=\overline{A}_{\mathbf{r}\mathbf{r}^+_{\nu}}-\overline{A}_{\mathbf{r},\mathbf{r}^{+}_{\mu}}$ for compactness.In the saddle point approximation, the functional integral in $Z$ will be dominated by the saddle point value of $\lambda$, which we then take out of the sum $\sum_{\mathbf{r}}$.We then apply the Fourier transform on the bosonic operator and obtain the bosonic spinon action 

\begin{equation}\label{bosonaction}
S_{b}=\int \frac{d\omega_n}{2\pi}\sum_{\mathbf{k}>0}\hat{b}^{\dag}_{\mathbf{k}}\left(M(\mathbf{k})+\frac{\omega^2_n}{2J_{zz}}I+\lambda\right)\hat{b}_{\mathbf{k}}
\end{equation}
where $\hat{b}_{\mathbf{k}}=(b^A_{\mathbf{k}},b^{A*}_{-\mathbf{k}},b^B_{\mathbf{k}},b^{B*}_{-\mathbf{k}})^T$ describing the bosonic spinons living at the sites of the dual diamond lattice, corresponding to the centers of `up' and `down' tetrahedra, defining the A and B sublattices respectively of the diamond lattice.The $4\times 4$ matrix $M$ given by
\begin{equation}
M(\mathbf{k})=\begin{pmatrix}
  M_{11}(\mathbf{k}) & 0 & 0 & 0 \\
  0 & M_{11}(\mathbf{k}) & 0 & 0 \\
  0  & 0  &M_{11}(\mathbf{k}) & 0  \\
  0 & 0 & 0 & M_{11}(-\mathbf{k})
 \end{pmatrix}
\end{equation}
where
\begin{equation}\label{A11k}
M_{11}(\mathbf{k})=-J_{\pm}\sum_{\mu\neq\nu}\langle s^{-}_{\mathbf{r}\mathbf{r}_{\mu}}\rangle\langle s^{+}_{\mathbf{r}\mathbf{r}_{\nu}}\rangle e^{-i\mathbf{k}\cdot(\mathbf{e}_{\mu}-\mathbf{e}_{\nu})}
\end{equation}
which is nicely diagonal for our QSL and QSSF states.The diagonal elements automatically are the four eigenvalues $\epsilon_m$, which consist of triplet degenerate eigenvalues ($\epsilon_{2,3,4}(\mathbf{k})=M_{11}(\mathbf{k})$) and one nondegenerate singlet state eigenvalue ($\epsilon_1(\mathbf{k})=M_{11}(-\mathbf{k})$).We have checked that $M_{11}(\mathbf{k}),M_{11}(-\mathbf{k})$ are real-valued, as should be the case for an energy eigenvalue.It is to be noted that the eigenvalue is independent of $J_{\pm\pm}$.This is valid so long as $J_{\pm\pm}<J^{\mathrm{crit}}_{\pm\pm}$ so that we are in either QSL or QSSF states.We find that, interestingly, while there is a gap between the singlet state and the three degenerate triplet states for the $U(1)$ QSL, the four states become degenerate in the QSSF state; that is, the singlet-triplet energy gap vanishes in the spin superfluid state.This is an important observation.The corresponding eigenstates are 4-component vectors with elements that can be written as $\phi^m_{\alpha}$ which represents the $\alpha^{\mathrm{th}}$-element of the $m^{\mathrm{th}}$ eigenstate.The diagonal-ness of the matrix $M$ greatly simplifies the calculations since $\phi^m_{\alpha}=\delta^m_{\alpha}$.In order to compute the variational energy later, a compute the single-particle Green's function of the bosonic spinon
\begin{equation}
G_{\alpha\beta}(\mathbf{k},\omega)=\sum_{m}\frac{{\phi^m_{\alpha}}^*\phi^m_{\beta}}{\frac{\omega^2}{2J_{zz}}+\lambda+\epsilon_m(\mathbf{k})}
\end{equation}
We then sum over Matsubara frequencies, which for $T=0$ becomes an integral over frequency, which can be done using contour integral, giving
\[
G_{\alpha\beta}(\mathbf{k})=\int \frac{d\omega}{2\pi}\sum_{m}\frac{{\phi^m_{\alpha}}^*\phi^m_{\beta}}{\frac{\omega^2}{2J_{zz}}+\lambda+\epsilon_m(\mathbf{k})}
\]
\begin{equation}\label{spinonGF}
=\sum_m\sqrt{\frac{J_{zz}}{2(\lambda+\epsilon_m(\mathbf{k}))}}{\phi^m_{\alpha}}^*\phi^m_{\beta}
\end{equation}

The variational energy of interest is given by
\begin{equation}\label{VEVlatticeQED}
\langle H_{\mathrm{QED}}\rangle=
\frac{J_{zz}}{2}\sum_{\mathbf{r}}\langle Q^2_{\mathbf{r}}\rangle -J_{\pm}\sum_{\mathbf{r}}\sum_{\mu\neq\nu}\langle b^{\dag}_{\mathbf{r}_{\mu}}b_{\mathbf{r}_{\nu}}\rangle\langle s^{-\eta_{\mathbf{r}}}_{\mathbf{r}\mathbf{r}_{\mu}}\rangle\langle s^{+\eta_{\mathbf{r}}}_{\mathbf{r}\mathbf{r}_{\nu}}\rangle
\end{equation}
The task now is to compute the expectation values $\langle Q^2_{\mathbf{r}}\rangle$ and $\langle b^{\dag}_{\mathbf{r}_{\mu}}b_{\mathbf{r}_{\nu}}\rangle$ in Eq.(\ref{VEVlatticeQED}) using the spinon single particle Green's function Eq.(\ref{spinonGF}).Note that the $\langle b^{\dag}_{\mathbf{r}_{\mu}}b_{\mathbf{r}_{\nu}}\rangle$ is a particle-hole order parameter between spinons on the same sublattice, and is nonzero even for the $U(1)$ QSL and the QSSF under consideration.To compute $\langle Q^2_{\mathbf{r}}\rangle$, we note that $b_{\mathbf{r}}$ and $Q_{\mathbf{r}}$ satisfy a commutation relation $[\varphi_{\mathbf{r}},Q_{\mathbf{r}}]=i$ where $b_{\mathbf{r}}=\exp(-i\varphi_{\mathbf{r}})$ satisfying $b^{\dag}_{\mathbf{r}}b_{\mathbf{r}}=1$.Due to this commutation relation, we can view $b_{\mathbf{r}}$ (via $\varphi_{\mathbf{r}}$) and $Q_{\mathbf{r}}$ as a pair of canonically conjugate position-momentum variables $x$ and $p$ respectively.To be precise, we write $\mathbf{Q}_{\mathbf{r}}=(Q^x_{\mathbf{r}},Q^y_{\mathbf{r}})$ and $b_{\mathbf{r}}=b^x_{\mathbf{r}}+ib^y_{\mathbf{r}}$.Then $p_{x(y)}=\sqrt{J_{zz}}Q^{x(y)}_{\mathbf{r}}$ and $x(y)=b_{\mathbf{r}}/\sqrt{J_{zz}}$.We then use a standard result from virial theorem or equipartition theorem; $\langle p^2\rangle=\langle \dot{x}^2\rangle$.As a consequence,
\[
\frac{J_{zz}}{2}\sum_{\mathbf{r}}\langle Q^2_{\mathbf{r}}\rangle=\frac{1}{2}\sum_{\mathbf{r}}\langle p^2_x\rangle+\langle p^2_y\rangle
=\frac{1}{2}\sum_{\mathbf{r}}\langle \dot{x}^2\rangle+\langle \dot{y}^2\rangle
\]
\[
=\sum_{\mathbf{r}}\frac{1}{2J_{zz}}\langle \dot{b}^{\dag}_{\mathbf{r}}\dot{b}_{\mathbf{r}}\rangle=\int \frac{d\omega}{2\pi}\sum_{\mathbf{k}\in A,B}\frac{\omega^2}{2J_{zz}}\langle b^{\dag}_{\mathbf{k}}b_{\mathbf{k}}\rangle
\]
\[
=\int \frac{d\omega}{2\pi}\sum_{\mathbf{k}\in A,B}\frac{\omega^2}{2J_{zz}}\left(G_{11}(\mathbf{k},\omega)+G_{33}(\mathbf{k},\omega)\right)
\]
\begin{equation}\label{chargeterm}
=\frac{1}{2}\sum_{\mathbf{k}}\sum_m\sqrt{2J_{zz}(\lambda+\epsilon_m(\mathbf{k}))}\left({\phi^m_{1}}^*\phi^m_{1}+{\phi^m_{3}}^*\phi^m_{3}\right)
\end{equation}

On the other hand, using Eqs.(\ref{A11k}) and (\ref{spinonGF}), we have 
\[
 -J_{\pm}\sum_{\mathbf{r}}\sum_{\mu\neq\nu}\langle b^{\dag}_{\mathbf{r}_{\mu}}b_{\mathbf{r}_{\nu}}\rangle\langle s^{-\eta_{\mathbf{r}}}_{\mathbf{r}\mathbf{r}_{\mu}}\rangle\langle s^{+\eta_{\mathbf{r}}}_{\mathbf{r}\mathbf{r}_{\nu}}\rangle=\sum_{\mathbf{k}}M_{11}(\mathbf{k})\langle b^{\dag}_{\mathbf{k}}b_{\mathbf{k}}\rangle 
\]
\begin{equation}\label{kineticterm}
=\sum_{\mathbf{k}}\sum_m\sqrt{\frac{J_{zz}}{\lambda+\epsilon_m(\mathbf{k})}}M_{11}(\mathbf{k})({\phi^m_1}^*\phi^m_1+{\phi^m_3}^*\phi^m_3) 
\end{equation}

Using $\phi^m_{\alpha}=\delta^m_{\alpha}$ and substituting it into Eqs.(\ref{chargeterm}) and (\ref{kineticterm}) in conjunction with Eq.(\ref{VEVlatticeQED}), we obtain
\begin{equation}
\langle H_{\mathrm{QED}}\rangle=\sum_{\mathbf{k},m=1,3}\frac{1}{2}\sqrt{2J_{zz}(\lambda+\epsilon_m(\mathbf{k}))}
+\frac{J_{zz}M_{11}(\mathbf{k})}{\sqrt{2J_{zz}(\lambda+\epsilon_m(\mathbf{k}))}}
\end{equation}

\begin{figure}
 \includegraphics[angle=0,origin=c, scale=0.40]{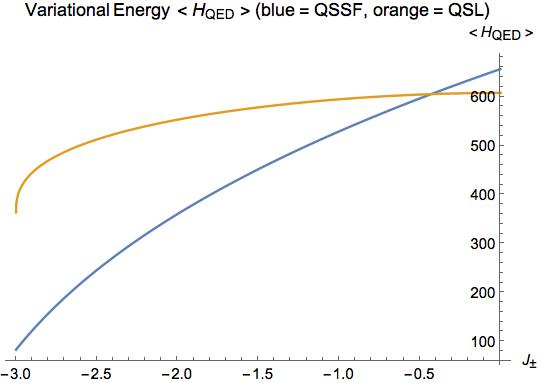}
 \label{fig:VariationalEnergyantiferromagnetic}
 \caption{
 The variational energy $\langle H_{\mathrm{QED}}\rangle$ of the $U(1)$ QSL and QSSF states vs. $J_{\pm}$ with $J_{zz}=1,J_{\pm\pm}<J^{\mathrm{crit}}_{\pm\pm},J_{z\pm}=0,\lambda_{\mathrm{QSL}}=3.0,\lambda_{\mathrm{QSSF}}=3.5$, and $J_{\pm}<0$ (antiferromagnetic).}
 \end{figure}
 \begin{figure}
 \includegraphics[angle=0,origin=c, scale=0.40]{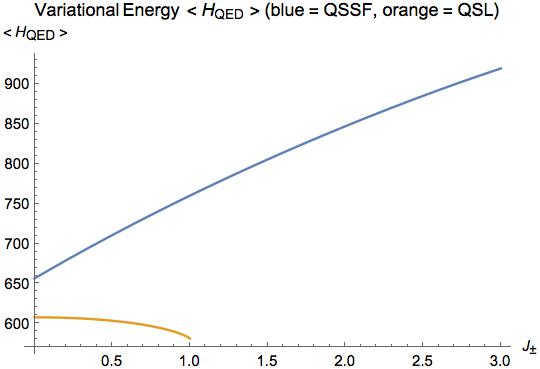}
 \label{fig:VariationalEnergy}
 \caption{
 The variational energy $\langle H_{\mathrm{QED}}\rangle$ of the $U(1)$ QSL and QSSF states vs. $J_{\pm}$ with $J_{zz}=1,J_{\pm\pm}<J^{\mathrm{crit}}_{\pm\pm},J_{z\pm}=0,\lambda_{\mathrm{QSL}}=3.0,\lambda_{\mathrm{QSSF}}=3.5$, and $J_{\pm}>0$ (ferromagnetic).}
 \end{figure}
 
The result for the variational energy $\langle H_{\mathrm{QED}}\rangle$ as function of $J_{\pm}$ at fixed $J_{zz},J_{\pm\pm}$ and $\lambda$ for the $U(1)$ QSL and the QSSF states using the ansatz given in Eq.(7) in the main text, applicable to the antiferromagnetic ($J_{\pm}<0$) case, is shown in Fig. (3).We clearly see that the quantum spin superfluid has lower energy than the $U(1)$ quantum spin liquid state for large enough $|J_{\pm}|$ in this antiferromagnetic $J_{\pm}<0$ case.The crossing point marks a quantum phase transition from $U(1)$ QSL state at small $|J_{\pm}|$ to QSSF state at larger $|J_{\pm}|$ for this antiferromagnetic ($J_{\pm}<0$) case.Crucially, when one repeats the calculations using the ferromagnetic ansatz $\overline{A}_{\mathbf{r},\mathbf{r}_{0}}=0,\overline{A}_{\mathbf{r},\mathbf{r}_{1}}=\pi, \overline{A}_{\mathbf{r},\mathbf{r}_{2}}=0,\overline{A}_{\mathbf{r},\mathbf{r}_{3}}=\pi$ corresponding to zero flux, one will find that the QSSF state has consistently higher energy than the QSL state, as shown in Fig. (4).This means the QSSF state does not occur for ferromagnetic case ($J_{\pm}>0$).The quantum spin superfluid state thus exists only in the antiferromagnetic quantum spin ice, in precise agreement with the mean-field anstaz and flux analysis presented in the main text.This result is the variational energy calculation basis for our proposed ground state phase diagram given in Fig. (2) in the main text with the quantum spin superfluid state in it.In this ground state phase diagram, the straight line of the phase boundary in the QSL-QSSF transition arises from the $J_{\pm\pm}$-independence of the variational energy $\langle H_{\mathrm{QED}}\rangle$ of both liquid states due to the absence of both pairing and sublattice-mixing particle-hole orders.However, the spin superfluid jump decreases as one increases $J_{\pm\pm}$.

As supplementary note, in obtaining Figs.(3) and (4), we have chosen $\lambda_{\mathrm{QSSF}}>\lambda_{\mathrm{QSL}}$ because the spinon condensation in QSSF state raises its Lagrange multiplier, according to $\lambda=\lambda_0+\lambda'/N^2_{\mathrm{u.c.}}$ where $\lambda_0$ is for the case without spinon condensation (as in QSL), $\lambda'$ is a coupling-dependent parameter, and $N_{\mathrm{u.c}}$ is the number of unit cells \cite{LeeOnodaBalentsS}.The $\lambda_0$ is determined by the minimum of the eigenvalue $\epsilon(\mathbf{k})$ which we find to give $\lambda_0=3$ in units where $J_{zz}=1$.But, the $\lambda'$ has to be determined from a self-consistent variational calculation on a finite system and $\lambda'$ then depends on the system size $N_{\mathrm{u.c}}$.The resulting critical $|J^c_{\pm}|\simeq 0.4 J_{zz}$ marking the QSL-QSSF transition as illustrated in Fig.(3) turns out to be of the same order of magnitude as that from numerical quantum Monte-Carlo result which gives $|J^c_{\pm}|\simeq 0.104 J_{zz}$ \cite{BojosenOnodaS}\cite{IsakovS}.


\begin{thebibliography}{1}
\bibitem{DeconfinedHiggs}The possibility of having a deconfined Higgs condensate of bosons coupled to gauge field is explained in \cite{SuppMat}.
\end{thebibliography}
\end{document}